\documentclass[preprint,12pt]{elsarticle}

\usepackage{graphicx}

\usepackage{amsmath,amssymb}

\usepackage{enumerate}

\journal{}

\begin{document}

\begin{frontmatter}



\title{Cyclic Calorons}



\author[Kit]{Atsushi Nakamula}
\ead{nakamula@sci.kitasato-u.ac.jp}
\address[Kit]{Department of Physics, School of Science, Kitasato University,
Sagamihara, Kanagawa 252-0373, Japan}

\author[TUS]{Nobuyuki Sawado}
\ead{sawado@ph.noda.tus.ac.jp}
\address[TUS]{Department of Physics, Tokyo University of Science,
 Noda, Chiba 278-8510, Japan}

\begin{abstract}
The Nahm data of periodic instantons, often called calorons, with spatial $C_N$-symmetries
are considered, by applying Sutcliffe's ansatz
for the monopoles with $C_N$-symmetries.
The bulk data of calorons are shown to enjoy the periodic Toda lattice,
and the solutions are given in terms of elliptic theta functions.
The case of $N=3$ calorons are investigated in detail.
It is found that the ``scale parameters" of these calorons have upper bounds 
in their values, so that they do not have the large scale, or monopole, limits.
The instanton limit of the $C_3$-symmetric caloron is  obtained.

\end{abstract}


\end{frontmatter}


\section{Introduction}
\label{intro}
There has long been interest in the topological objects in field theories 
\cite{ManSut,Radu,Chamon,Zahed,Shnir,AH}.
In particular, the instanton solutions to anti-self-dual (ASD) Yang-Mills equations in $\mathbb{R}^4$
\cite{BPST},
as well as the Bogomoln'yi-Prasad-Sommerfield (BPS) monopole solutions to 
the Bogomoln'yi equations on $\mathbb{R}^3$ \cite{Bogo,PS}
are both attracted much attention in wide area of mathematical physics.
The instantons on partially compactified space $\mathbb{R}^3\times S^1$ are called calorons
\cite{HS},
which are interpreted as the instantons of finite temperature gauge theories.
The fascinating feature of calorons is their non-trivial holonomy around $S^1$ at the spatial infinity
\cite{LeeLu,KvB,BNvB,Harland,NakaSaka}.
The appearance of the non-trivial holonomy 
is considered as giving a criterion of the confinement in pure Yang-Mills gauge theories
\cite{DiaGro, GroSli,Sli}.

Despite the extreme interest for such topological objects, a few
analytic descriptions are known for instantons, monopoles and calorons.
For the construction of these solutions, the ADHM/Nahm formalism is familiar.
In the formalism, instead of solving the original partial differential equations,
what we have to find first is the ADHM/Nahm data, which are solutions to algebraic equations and/or 
ordinary differential equations; thereafter we make a transformation into the ``real" gauge fields.
However, even for the ADHM/Nahm data, not so many exact, or analytic, solutions are known explicitly.
In general, if we need the analytic ADHM/Nahm data, we have to impose the system an appropriate symmetry
by ansatz.

In this paper, we construct the Nahm data of calorons associated with instanton charge $N$
which have $N$-th rotational symmetries around an axis ($C_N$-symmetries).
We will apply the $C_N$-symmetric ansatz for monopole Nahm data given by Sutcliffe 
\cite{SutPLB,SutNP}
as the bulk Nahm data of calorons.
The ansatz makes the Nahm equations into the celebrated periodic Toda lattice equations.
Although the uniqueness of the $C_N$-symmetric ansatz for monopoles was proved by Braden \cite{Br},  
the analytic Nahm data compatible with appropriate monopole boundary conditions was
 not known in analytic form,
except for the higher symmetric case such as Platonic symmetries \cite{HMM}.
It is expected that the monopole Nahm data of charge $N$ is solved by genus $N-1$ abelian integral
 in general \cite{SutPLB,ES}.
However, we claim here that there can be  the ``bulk" Nahm data of $C_N$-symmetric calorons
 in terms of elliptic, or Jacobi, theta functions.
As will be seen in the following sections,
 the distinct criterion on the boundary conditions of the Nahm data 
between monopoles and calorons enables to determine the analytic Nahm data with $C_N$-symmetries. 
As a remarkable result, the caloron Nahm data obtained here do not have the monopole,
 or large scale, limits.
The large period, or instanton, limit of the $C_N$-symmetric calorons is also investigated.
We restrict ourselves to consider the calorons of the $SU(2)$ gauge theory, and
of trivial holonomy cases, the generalization to the higher rank gauge groups, and
the non-trivial holonomy ones are given in elsewhere.
We will illustrate the construction with the $C_3$-symmetric Nahm data, as an introductory work.

The  paper is organized as follows.
In section 2, beginning with a short review on the symmetric Nahm data,
we consider the $C_N$-symmetric Nahm data of  calorons by applying Sutcliffe's ansatz.
The reduced Nahm equations are solved in terms of elliptic theta functions.
In section 3, we obtain the exact form of the Nahm data with $C_3$-symmetry, as a first example.
Section 4 is devoted to conclusion and discussion.



\section{The Nahm data of calorons with $C_N$-symmetries}

\subsection{Symmetric Nahm data}
In this section, we consider  the Nahm data of $SU(2)$ calorons  of
charge $N$ with $C_N$-symmetries.
The Nahm data of calorons are composed of two parts, the bulk data $T_j(s) \;(j=1,2,3),$ and the
boundary data $W$ \cite{Nahm}.
The former is three $N\times N$ matrix valued regular functions periodic in $s$; 
we take the fundamental period $[-\mu,\mu]$ here.
Throughout this paper, we take the gauge $T_0(s)=0$.
The boundary data is an $N$-dimensional row vector of quaternion entries.
In the Nahm construction of calorons, the Nahm data $\{T_j(s),W\}$ are defined by
the following four conditions.
The first is the differential equations for the bulk data, called Nahm equations,
\begin{equation}
T'_j(s)=\frac{i}{2}\epsilon_{jkl}[T_j(s),T_k(s)],\label{Nahm eq}
\end{equation}
where $i,j$ and $k$ run through $1$ to $3$, $\epsilon_{ijk}$ is totally anti-symmetric tensor, 
and the derivative is with respect to the variable $s$.
The second condition is the hermiticity for the bulk data 
\begin{equation}
T_j^\dag(s)=T_j(s),\label{hermiticity}
\end{equation}
and the third ones are the so called reality conditions
\begin{equation}
T_j^t(s)=T_j(-s).\label{reality}
\end{equation}
The fourth conditions are the relations between the bulk data and the boundary data,
 called the matching conditions
\begin{equation}
T_j(-\mu)-T_j(\mu)=\frac{1}{2}\mathrm{tr}_2\,\sigma_jW^\dag W,\label{matching}
\end{equation}
where the trace is taken for the quaternions.
The $SU(2)$ caloron gauge fields can be obtained by the Nahm data $\{T_j(s),W\}$
through the Nahm transform.

Next we consider the symmetry of caloron Nahm data under the action of $SO(3)$,
a rotation in the configuration space \cite{Ward}.
Let us denote $R$ an element of the subgroup of $SO(3)$, \textit{i.e.},
for a spatial rotation of a position vector we have $x_j\mapsto x'_j=R_{jk}x_k$,
where $R_{jk}$ is an image of $R$ in the 3-dimentional orthogonal representation
of $SO(3)$.
We also denote $R_2$ the image of $R$ in the $2$-dimensional irreducible 
representation of $SU(2)$, which gives rise to a spatial rotation of
quaternions, $x\mapsto x'=R_2xR_2^{-1}$,
where $x=x_\mu e_\mu$ and the quaternion basis are given by
$e_\mu=(1,-i\sigma_1,-i\sigma_2,-i\sigma_3)$.

A caloron Nahm data is said to be symmetric under the action of $R$, if the
Nahm data $\{T_j,W\}$ enjoy
\begin{equation}
R_N T_jR_N^{-1}=R_{jk}T_k,\label{bulkrotation}
\end{equation}
and
\begin{equation}
R_N\otimes R_2\,W^\dag=W^\dag \hat q,\label{boundaryrotation}
\end{equation}
where $R_N$ denotes the image of $R$ in $SL(n,\mathbb{C})$,
and $\hat{q}$ is a unit quaternion, \textit{i.e.}, $\hat{q}^\dag \hat{q}=1$.
For the case of $C_N$-symmetris around the ``3"-axis, we may have a choice
\begin{equation}
R_N=\omega^l \;\mathrm{diag.}[\omega^{N-1},\omega^{N-2},\dots,\omega,1],\label{R_N}
\end{equation}
where $\omega$ is an $N$-th root of unity and $l=0,1,\dots,N-1$ \cite{Br}.


\subsection{$C_N$-symmetries and Hermiticity of the bulk data}

Let us now consider the bulk Nahm data satisfying the $C_N$-symmetric condition (\ref{bulkrotation}).
For this purpose, we will apply the ansatz for
 the monopole Nahm data with $C_N$-symmetries given by Sutcliffe over a decade ago \cite{SutPLB,SutNP}.
The form of the $N\times N$ bulk data is given in terms of differentiable functions 
$f_j(s)$ and $p_j(s)\;(j=0,1,2\dots,N-1)$ as
\begin{align}
&T_1=\frac{1}{2}\left[
	\begin{array}{ccccccc}
	& f_{1}  &            &         &  f_{0}   \\
f_{1}  &        & f_2        &         &         \\
	& f_2   &             & \ddots&       \\
     	&        & \ddots  &           & f_{N-1} \\
f_{0}  &       &             & f_{N-1}  &   \\
	\end{array}
\right],\label{ansatz T_1}\\
&T_2=
\frac{i}{2}\left[
	\begin{array}{ccccccc}
	& -f_{1}  &            &         &  f_{0}   \\
f_{1}  &        & -f_2        &         &         \\
	& f_2   &             & \ddots&       \\
     	&        & \ddots  &           & -f_{N-1} \\
-f_{0}  &       &             & f_{N-1}  &   \\
	\end{array}
\right],\label{ansatz T_2}\\
&T_3=\frac{1}{2}\;\mathrm{diag.}\;[p_1,p_2,\cdots,p_{N-1},p_0],\label{ansatz T_3}
\end{align}
where we have omitted the argument of the functions.
From this ansatz, the hermiticity conditions are enjoyed if $f_j, p_j\in \mathbb{R}$.
Substituting the ansatz (\ref{ansatz T_1}--\ref{ansatz T_3}) into the Nahm equations (\ref{Nahm eq}),
we obtain the differential equations for $f_j(s)$'s and $p_j(s)$'s
\begin{align}
&f'_{j}=\frac{1}{2}\,f_{j}\,( p_{j+1}-p_{j})\label{equations f}\\
&p'_j=f_{j-1}^2-f_{j}^2,\label{equations p}
\end{align}
where the periodicity $f_{j+N}=f_j$ and $p_{j+N}=p_j$ is taking into account.
The system of differential equations (\ref{equations f},\ref{equations p}) is well known as the periodic 
Toda lattice.
Note that it is necessary, due to the Nahm equation, $\mathrm{tr}\,T'_3=\frac{1}{2} \sum_{j=0}^{N-1}p'_j=0$, 
which will be confirmed later.
Eliminating the $p_j(s)$'s in (\ref{equations f},\ref{equations p}), we have the equations for $f_j$'s,
\begin{equation} 
\frac{d^2}{ds^2}\log f_j^2=-f_{j+1}^2+2f_j^2-f_{j-1}^2,\label{equations f^2}
\end{equation}
which are well known form of Toda lattice except for the sign in the right hand side.

Having obtained the reduced Nahm equations for the bulk data,
we are in a position to find special solutions to (\ref{equations f^2}) appropriate for the
caloron Nahm data.
By introducing $\tau$-functions $\tau_j:=\tau(s,j),\;(j=0,1,\dots,N-1)$
\begin{align}
f_{j}^2&=-C^2\frac{\tau_{j-1}\tau_{j+1}}{\tau_j^2},\label{f_j}
\end{align}
the differential equations for $\tau_j$'s read from (\ref{equations f^2}) are
\begin{equation}
\frac{d^2}{ds^2}\log\tau_j=C^2\frac{\tau_{j-1}\tau_{j+1}}{\tau_j^2},\label{equations tau}
\end{equation}
where $C$ is a constant defined below.
Simultaneously, we find the expression for $p_j$'s by the $\tau$-functions 
from (\ref{equations f}), which is
\begin{align}
p_j&=\frac{d}{ds}\left(\log\frac{\tau_{j}}{\tau_{j-1}}\right).\label{p_j}
\end{align}
We now assume the following form to the $\tau$-functions in terms of elliptic, or Jacobi, theta functions
$\vartheta_\nu(u,q)$, where $u\in\mathbb{C}$ and $\nu=0,1,2$ or $3$,
and $q$ is the modulus parameter
\footnote{$\vartheta_0$ is also expressed as $\vartheta_4$ in the literatures.}.
The definition of these elliptic theta functions are given in Appendix.
The ansatz  is
\begin{equation}
\tau(s,j)=\exp\left(\frac{1}{2}\tilde{A} s^2+bs+\tilde{b} j\right)
\vartheta_\nu(\pm s+\kappa j+a,q),\label{ansatz}
\end{equation}
where $\tilde{A}, b, \tilde{b}, \kappa$ and $a$ are constants.
This form of the $\tau$-functions for the periodic Toda lattice was originally
introduced by M.Toda in 1967 \cite{Toda1}.
Substituting (\ref{ansatz}) into (\ref{equations tau}), 
we find a differential equation for the theta functions, 
\begin{equation}
A+C^{-2}\left(\log\vartheta_\nu(s_j)\right)''
=\frac{\vartheta_\nu(s_j-\kappa)\vartheta_\nu(s_j+\kappa)}{\vartheta_\nu^2(s_j)},
\label{diffeq for theta}
\end{equation}
where $s_j:=\pm s+\kappa j+a$, $A:=\tilde{A}C^{-2}$, and we have omitted the modulus dependence.
The differential equation (\ref{diffeq for theta}) is solved if the constants
are given by the special values of theta functions \cite{Toda,Toda1},
\begin{align}
&C^{-2}=\left(\frac{\vartheta_1(\kappa)}{\vartheta_1'(0)}\right)^2,\label{constant C}\\
&A=\tilde{A}C^{-2}=\left(
\frac{\vartheta_0(\kappa)}{\vartheta_0(0)}\right)^2-\frac{\vartheta_0''(0)}{\vartheta_0(0)}
\left(\frac{\vartheta_1(\kappa)}{\vartheta_1'(0)}\right)^2.\label{constant A}
\end{align}
Thus we have found the elliptic theta function solution to the Nahm equations (\ref{equations f})
and (\ref{equations p}),
\begin{align}
f_{j}^2=-C^2\frac{\vartheta_{\nu}(s_{j-1})\vartheta_{\nu}(s_{j+1})}{\vartheta_{\nu}(s_{j})^2},
\label{solution f^2}\\
p_j=\frac{d}{ds}\left(\log\frac{\vartheta_{\nu}(s_{j})}{\vartheta_{\nu}(s_{j-1})}\right).
\label{solution p}
\end{align}
Note that the exponential factors are not appeared in the expression.
By taking the square root of (\ref{solution f^2}), $f_j$'s are rewritten as
\begin{align}
f_{j}=\pm iC\frac{\sqrt{\vartheta_{\nu}(s_{j-1})\vartheta_{\nu}(s_{j+1})}}{\vartheta_{\nu}(s_{j})}.
\label{solution f}
\end{align}
We can take the plus sign in (\ref{solution f}) without loss of generality.

We now recall the periodicity of the theta functions $\vartheta_\nu(u+1)=\pm\vartheta_\nu(u)$, 
where the sign is depending on $\nu$, see Appendix.
Taking into account the definition $s_j:=\pm s+\kappa j+a$,
 we easily find that the periodicity $f_{j+N}=f_j$ and $p_{j+N}=p_j$
holds if we take $\kappa=1/N$.

Having obtained the special solutions to the Nahm equations, let us now fix 
the other conditions for the bulk Nahm data.
For the hermiticity (\ref{hermiticity}), it is necessary that $f_j, p_j\in\mathbb{R}$ as mentioned earlier. 
From (\ref{solution p}) and (\ref{solution f}), it is sufficient to take $\vartheta_{\nu}(s_{j})$ 
be real and positive valued on the region $s\in[-\mu,\mu]$,
 and $C\in i\mathbb{R}$, \textit{i.e.}, pure imaginary valued.
For these to be enjoyed, we choose the modulus parameter $q$ takes real values $0<q<1$,
then we find $\vartheta_\nu(s_{j})$ is positive real valued on $s\in[-\mu,\mu]$ for
$\nu=0$ and $3$.
On the other hand, for $\nu=1$ or $2$, $\vartheta_\nu(s_{j})$ has a zero
and then changes sign on the real axis, so that
the number in the square root of (\ref{solution f}) will be negative.
One can see there does not exist a case in which all of the $f_j$'s take real valued 
simultaneously, for $\nu=1$ or $2$.
Hence we eliminate the solution of $\nu=1$ and $2$,
 and concentrate on the solutions of $\nu=0$ or $3$, hereafter.
Next we fix the constant $C$ to be pure imaginary.
From (\ref{constant C}), we find
\begin{align}
C=\pm \frac{\vartheta_1'(0)}{\vartheta_1(\kappa)},\label{branch C}
\end{align}
where both the numerator and the denominator have a factor $q^{1/4}$,
\begin{align}
&\vartheta_1'(0)=2\pi q^{1/4}\prod_{m=1}^\infty(1-q^{2m})^3,\\
&\vartheta_1(\kappa)=2q^{1/4}\sin\kappa\pi\prod_{m=1}^\infty(1-2q^{2m}\cos2\kappa\pi+q^{4m}).
\end{align}
Hence, it is satisfactory if we take the relative branch of $q^{1/4}$ 
in (\ref{branch C}) so that $C\in i\mathbb{R}$.

\subsection{The reality conditions}

We now consider the reality conditions (\ref{reality}), \textit{i.e.}, $T_j^t(s)=T_j(-s)$.
One can easily find that the bulk Nahm data obtained above do not satisfy the conditions.
To illustrate, for the diagonal matrix $T_3(s)$, the reality condition reads all of  
the components $p_j(s)$ have to be even functions, however it
is not the case for (\ref{solution p}) in general.
Therefore, we have to show that there is another basis of the bulk Nahm data
 in which the reality conditions are apparent, by a unitary transformation $U_N$,
\begin{equation}
T_j\mapsto\tilde{T}_j=U_NT_jU_N^{-1},
\end{equation}
where
\begin{equation}
\tilde{T}_j^t(s)=\tilde{T}_j(-s).
\end{equation}
In the new basis of the bulk data $\tilde{T}_j$, however, the $C_N$-symmetries are
not apparent.
The situation is similar to the monopoles with Platonic symmetries \cite{HMM}.
We hereafter refer the primary $T_j$ as the bulk data in ``$C_N$-symmetric basis", and
$\tilde{T}_j$ as that in ``reality basis". 
In the next section, we show the explicit form of the unitary transformation and 
confirm there exists a reality basis  for $N=3$.
Finally, the transformation of the boundary data
from the $C_N$-symmetric basis to the reality basis can be given by (\ref{matching}) as 
\begin{equation}
\tilde{W}^\dag=U_NW^\dag,\ \tilde{W}=WU_N^{-1},\label{boundarytransform}
\end{equation}
where we define $\tilde W$ as the boundary data in the reality basis.

\section{Example: $C_3$-symmetric calorons}
\subsection{The bulk data}
Let us now illustrate the Nahm data of the $C_N$-symmetric calorons by considering the most 
simple case, $N=3$.
The case of larger values of $N$ will be considered in forthcoming articles.
In general, there will be an individual facet in the construction for $N$ by $N$, as in the 
cases of Platonic monopoles \cite{HMM}.

First we find the bulk Nahm data in the $C_3$-symmetric basis. 
The ansatz for $N=3$ reads from (\ref{ansatz T_1})--(\ref{ansatz T_3}) 
\begin{align}
&T_1=\frac{1}{2}\left[
	\begin{array}{ccc}
		0	    & f_{1}     &   f_{0}   \\
         f_{1}           & 0           &   f_{2} \\
	f_{0}    &    f_{2}  & 0  \\
	\end{array}
\right],\\
&T_2=\frac{i}{2}\left[
	\begin{array}{ccc}
		0	    & -f_{1}     &   f_{0}   \\
         f_{1}           & 0           &   -f_{2} \\
	-f_{0}    &    f_{2}  & 0  \\
	\end{array}
\right]\\
&T_3=\frac{1}{2}\;\mbox{diag.}\;[p_1,p_2,p_0].
\end{align}
The theta function solutions are given by (\ref{solution f^2}) and (\ref{solution p}) with
periodicity $f_{j+3}=f_j$ and $p_{j+3}=p_j$.
Hence we choose $\kappa=1/3$ and $a=0$ for $s_j=\pm s+\kappa j+a$ so that
$\vartheta_\nu(s_{j+3})=\vartheta_\nu(s_{j})$.
The ambiguity of the sign in $s_j$ is fixed to be consistent with the matching conditions (\ref{matching}),
which will be considered later.
We find the solutions are 
\begin{align}
f_0(s)=iC\frac{\sqrt{\vartheta_\nu(s_{1})\vartheta_\nu(s_{2})}}{\vartheta_\nu(s_0)},\label{f_0}\\
f_1(s)=iC\frac{\sqrt{\vartheta_\nu(s_{2})\vartheta_\nu(s_{0})}}{\vartheta_\nu(s_1)},\label{f_1}\\
f_2(s)=iC\frac{\sqrt{\vartheta_\nu(s_{0})\vartheta_\nu(s_{1})}}{\vartheta_\nu(s_{2})},\label{f_2}
\end{align}
and
\begin{align}
p_0(s)=\frac{d}{ds}\log\frac{\vartheta_\nu(s_{0})}{\vartheta_\nu(s_{2})},\label{p_0}\\
p_1(s)=\frac{d}{ds}\log\frac{\vartheta_\nu(s_1)}{\vartheta_\nu(s_{0})},\label{p_1}\\
p_2(s)=\frac{d}{ds}\log\frac{\vartheta_\nu(s_{2})}{\vartheta_\nu(s_{1})},\label{p_2}
\end{align}
where  $\nu=0$ or $3$, and $C=\vartheta'_1(0)/\vartheta_1(\kappa)\in i\mathbb{R}$,
which ensures the hermiticity.
In addition, we observe the consistency condition
\begin{equation}
\sum_{j=0}^2p_j(s)=\frac{d}{ds}\log\frac{\vartheta_\nu(s_{0})}{\vartheta_\nu(s_{2})}
\frac{\vartheta_\nu(s_{1})}{\vartheta_\nu(s_{0})}\frac{\vartheta_\nu(s_{2})}{\vartheta_\nu(s_{1})}=0,
\label{sumofp}
\end{equation}
holds. 

The next task is to fix the boundary data $W$.
For this to be done,
it is appropriate to carry out in the reality basis rather than the $C_3$-symmetric basis
obtained above.
Under the inversion $s\to-s$, we notice that 
\begin{align}
\vartheta_\nu(s_j)=\vartheta_\nu(\pm s+j/3)\longrightarrow
\vartheta_\nu(\mp s+ j/3)=\vartheta_\nu(\pm s-j/3)=\vartheta_\nu(s_{-j}).
\end{align}
Then we observe $f_0(s)$ and $p_2(s)$ of the $C_3$-symmetric bulk data
are even functions, on the other hand,
$f_1(s)$ and $f_2(s)$, and $p_0(s)$ and $p_1(s)$ swap each other.
Taking account of this fact, we suppose the following unitary transformation.
Defining a matrix
\begin{equation}
J_{3}=\left[
	\begin{array}{ccc}
0 & 0  & 1 \\
0 & 1  & 0 \\
1 & 0  & 0 \\
	\end{array}
\right],
\end{equation}
then we find 
\begin{equation}
U_{3}=\frac{1}{\sqrt{2}}(1_3+iJ_3)
=\frac{1}{\sqrt{2}}\left[
	\begin{array}{ccc}
   1       &0  &        i      \\
0 & 1+i      & 0               \\
    i 	 &0  &  1                   \\
	\end{array}
\right],
\end{equation}
is a unitary matrix.
The bulk data in the reality basis read
\begin{equation}
\tilde{T}_j=U_3\, T_j \,U_3^{-1},
\end{equation}
whose components are
\begin{align}
&\tilde{T}_1=\frac{1}{2}\left[
\begin{array}{ccc}
  0 	 & f_{+}-if_-  &   f_0   \\
 f_{+}+if_-  & 0      &  f_{+}-if_- \\
 f_0    & f_{+}+if_-  & 0  \\
	\end{array}
\right],\label{real T1}\\
\nonumber\\
&\tilde{T}_2=\frac{1}{2}\left[
\begin{array}{ccc}
f_0 	 & -f_{+}-if_-  &  0   \\
-f_{+}+if_-  & 0      &  f_{+}+if_- \\
	0   & f_{+}-if_-  & -f_0  \\
	\end{array}
\right],\label{real T2}\\
\nonumber\\
&\tilde{T}_3=\frac{1}{4}\left[
	\begin{array}{ccc}
-p_2	    & 0     &  i(p_0-p_1)   \\
    0    & 2p_2     &  0 \\
-i(p_0-p_1)    &    0  & -p_2  \\
	\end{array}
\right],\label{real T3}
\end{align}
where $f_\pm (s)=(f_1(s)\pm f_2(s))/2$.
From the behavior of $f_j$ and $p_j$ under the inversion $s\to-s$,
we find that $f_0, f_+$ and $p_2$ are even functions, and 
$f_-$ and $p_0-p_1$ are odd functions, respectively.
Thus, we can observe  that the reality conditions (\ref{reality}) hold for $\tilde{T}_j$'s.
Figure 1 and Figure 2 show the profile of the bulk data elements for the $\vartheta_0$ solution
and the $\vartheta_3$ solution, respectively.

\bigskip
\begin{center}
\includegraphics[height=27mm]{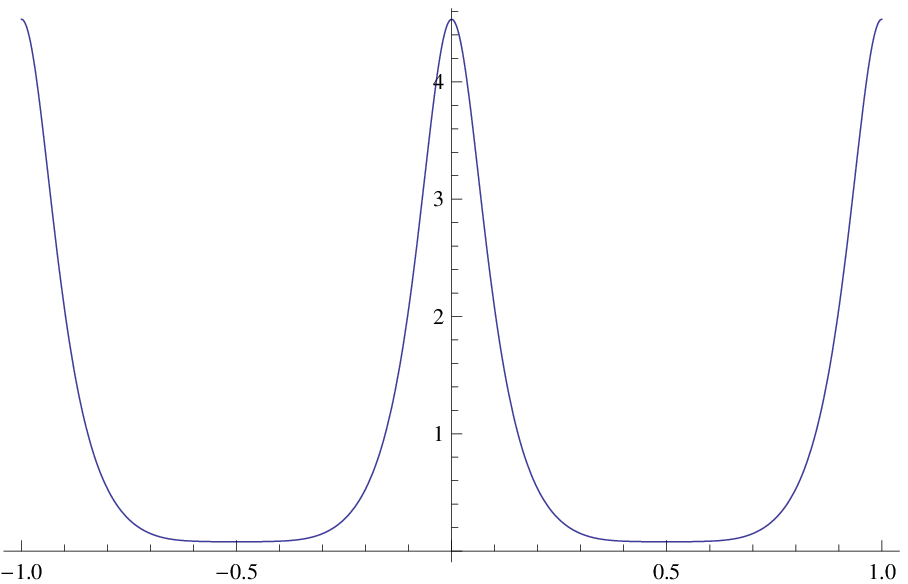}
$f_0(u)$
\hspace{1cm}
\includegraphics[height=27mm]{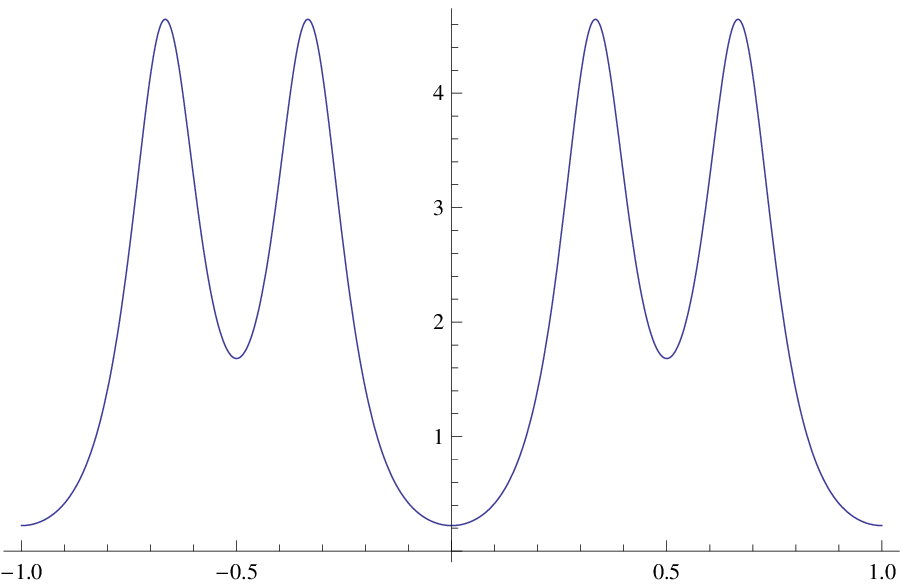}
$f_+(u)$

\bigskip
\includegraphics[height=27mm]{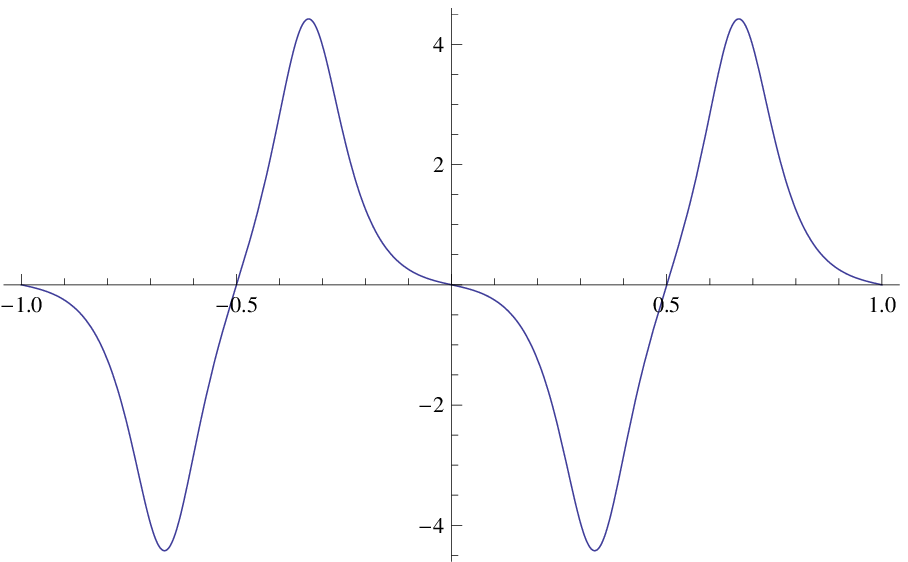}
$f_-(u)$

\bigskip
\includegraphics[height=27mm]{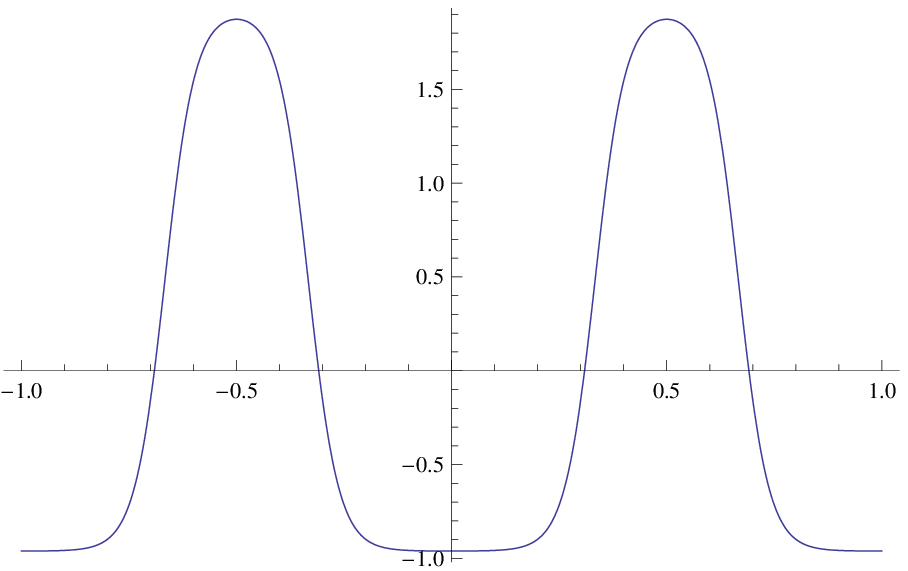}
$p_2(u)$
\hspace{0.5cm}
\includegraphics[height=27mm]{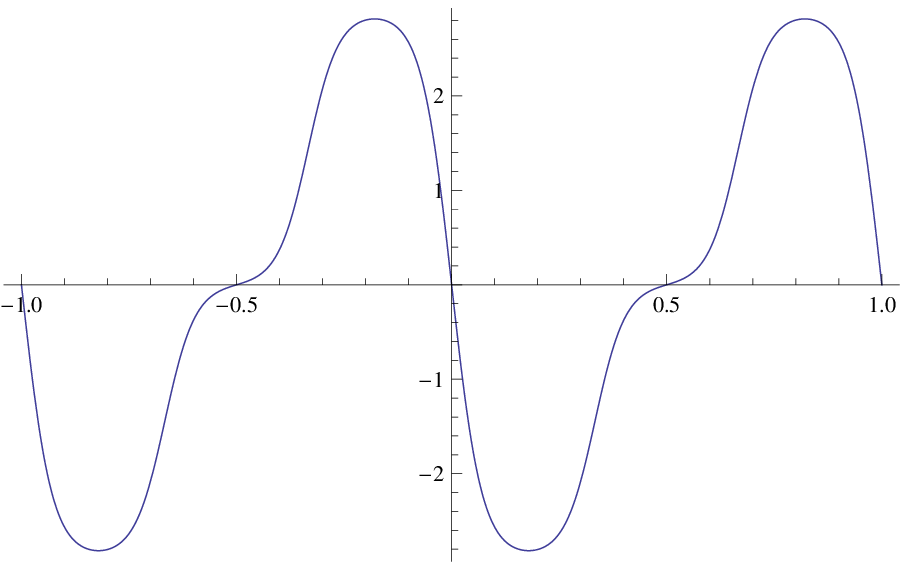}
$p_0(u)-p_1(u)$

\bigskip
Figure 1. \\
The profiles of the bulk data for the $\vartheta_0$ solution.
\end{center}

\bigskip

\begin{center}
\includegraphics[height=27mm]{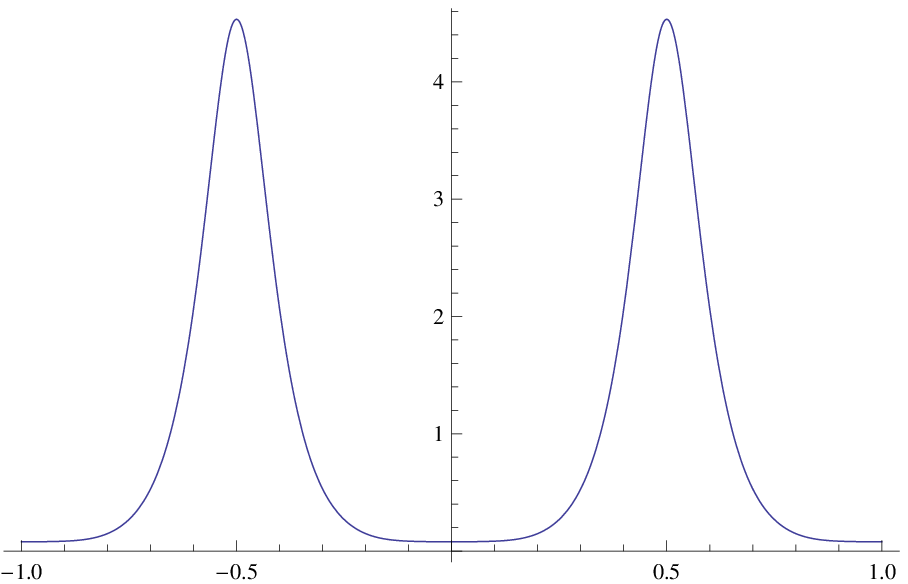}
$f_0(u)$
\hspace{1cm}
\includegraphics[height=27mm]{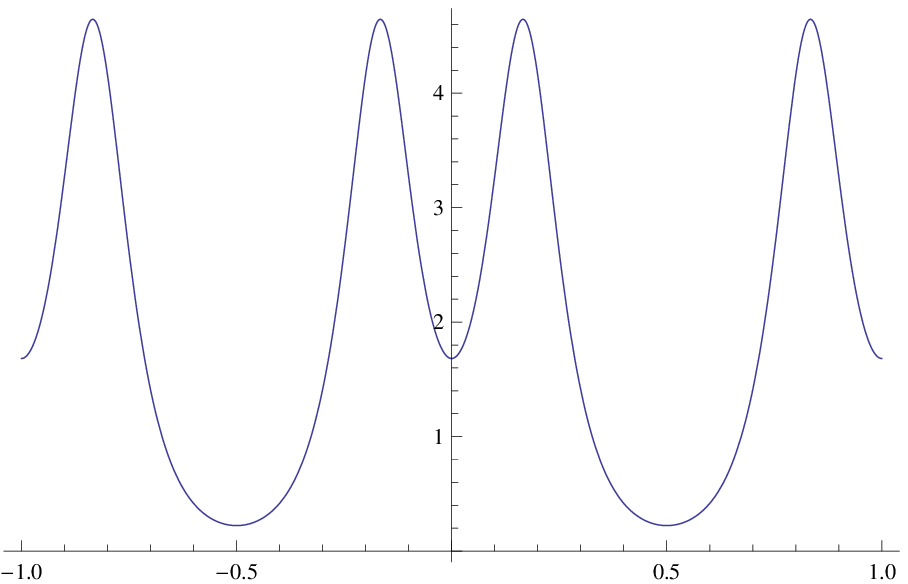}
$f_+(u)$

\bigskip
\includegraphics[height=27mm]{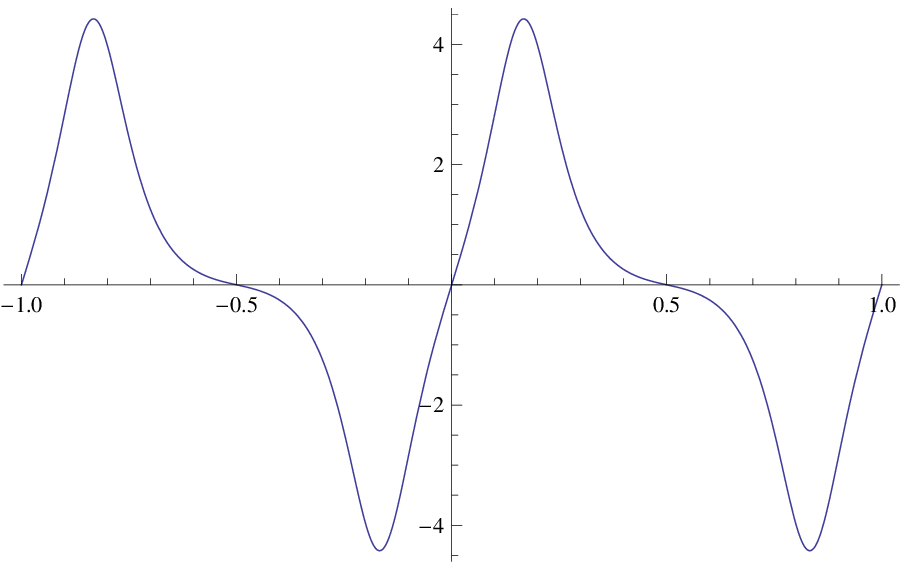}
$f_-(u)$

\bigskip
\includegraphics[height=27mm]{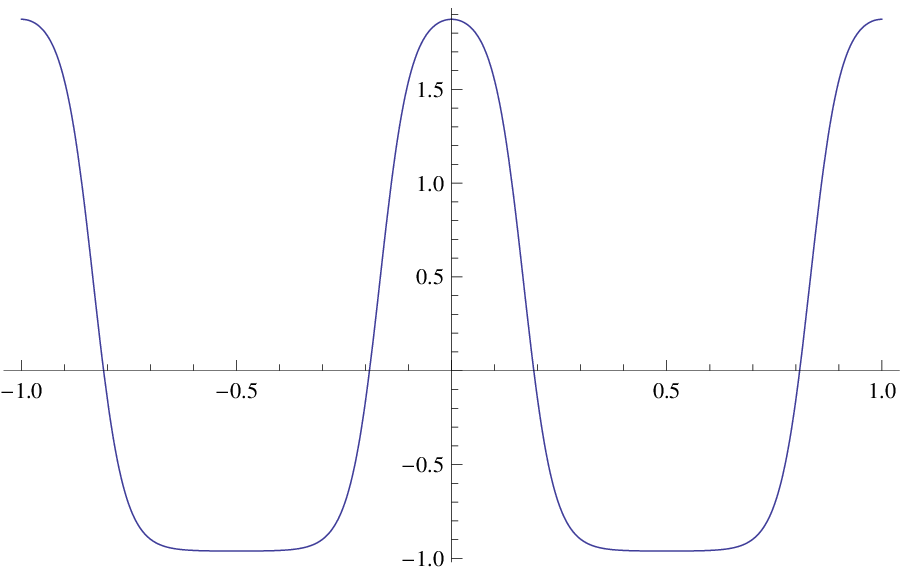}
$p_2(u)$
\hspace{0.5cm}
\includegraphics[height=27mm]{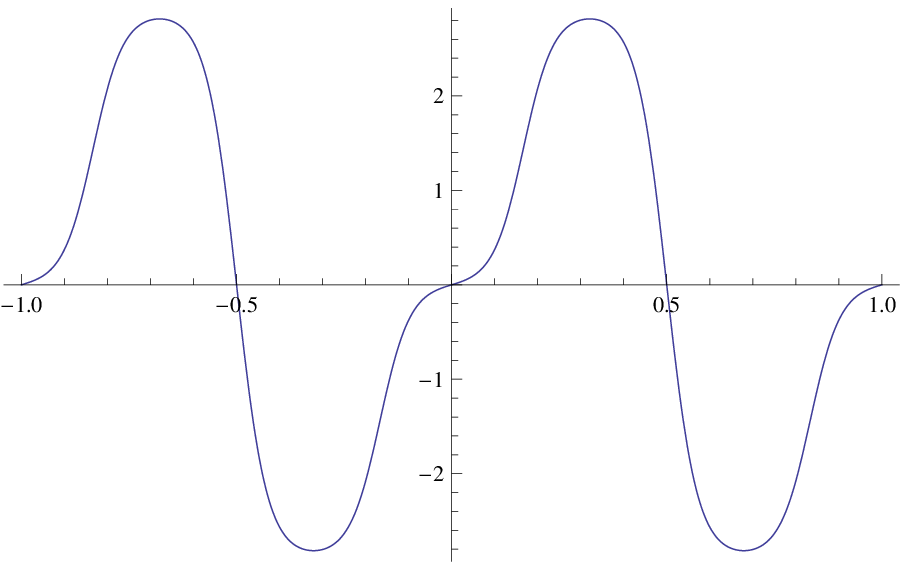}
$p_0(u)-p_1(u)$

\bigskip
Figure 2. \\
The profiles of the bulk data for the $\vartheta_3$ solution.
\end{center}

For larger values of $N$,  one can similarly define the unitary matrices which transform
the $C_N$-symmetric basis into the reality basis. 
The detail will be given in the forthcoming articles.

\subsection{The boundary data}
The next step is to find the exact form of the boundary data in the reality basis,
\begin{equation}
\tilde{W}=(\lambda,\rho,\chi),
\end{equation}
where $\lambda,\;\rho$ and $\chi$ are quaternions with components 
$\lambda=(\lambda_0,\lambda_1,\lambda_2,\lambda_3)$, \textit{etc.}.
The right hand side of (\ref{matching}) turns out to be
\begin{align}
\frac{1}{2}\mathrm{tr}_2\,\sigma_j\tilde{W}^\dag \tilde{W}=&\frac{1}{2}\mathrm{tr}_2\,\sigma_j
\left[\begin{array}{ccc}
\lambda^\dag\lambda & \lambda^\dag\rho & \lambda^\dag\chi   \\
\rho^\dag\lambda & \rho^\dag\rho & \rho^\dag\chi   \\
\chi^\dag\lambda & \chi^\dag\rho & \chi^\dag\chi   \\
	\end{array}
\right]\nonumber\\
=&\frac{1}{2}\mathrm{tr}_2\,\sigma_j
\left[\begin{array}{ccc}
0 & \lambda^\dag\rho & \lambda^\dag\chi   \\
\rho^\dag\lambda & 0 & \rho^\dag\chi   \\
\chi^\dag\lambda & \chi^\dag\rho & 0   \\
	\end{array}
\right],\label{boundary-3}
\end{align}
where we have used $\lambda^\dag\lambda=\sum_{\mu=0}^3\lambda_\mu^2\in\mathbb{R}$ \textit{etc.}. 
Defining
\begin{align}
g(\mu)&:=\frac{1}{2}(f_-(-\mu)-f_-(\mu))=-f_-(\mu),\label{defofg}\\
h(\mu)&:=\frac{1}{4}\left\{(p_0(-\mu)-p_1(-\mu))-(p_0(\mu)-p_1(\mu))\right\}\nonumber\\
&=-\frac{1}{2}\left(p_0(\mu)-p_1(\mu)\right),
\label{defofh}
\end{align}
we find that the left hand side of (\ref{matching}) for $j=1$ is 
\begin{align}
&\tilde{T}_1(-\mu)-\tilde{T}_1(\mu)\nonumber\\
=&
\frac{1}{2}\left[\begin{array}{ccc}
0 & -if_-(-\mu) & 0   \\
if_-(-\mu) & 0 & -if_-(-\mu)   \\
0 & if_-(-\mu) & 0   \\
	\end{array}
\right]
-\frac{1}{2}\left[\begin{array}{ccc}
0 & -if_-(\mu) & 0   \\
if_-(\mu) & 0 & -if_-(\mu)   \\
0 & if_-(\mu) & 0   \\
	\end{array}
\right]
\nonumber\\
=&\left[\begin{array}{ccc}
0 & -ig(\mu) & 0   \\
ig(\mu) & 0 & -ig(\mu)   \\
0 & ig(\mu) & 0   \\
	\end{array}
\right].\label{boundary-T1-3}
\end{align}
Similarly, we find the cases for $j=2$ and $3$ 
\begin{align}
\tilde{T}_2(-\mu)-\tilde{T}_2(\mu)
=&\left[\begin{array}{ccc}
0 & -ig(\mu) & 0   \\
ig(\mu) & 0 & ig(\mu)   \\
0 & -ig(\mu) & 0   \\
	\end{array}
\right],\label{boundary-T2-3}\\
\tilde{T}_3(-\mu)-\tilde{T}_3(\mu)=&
\left[\begin{array}{ccc}
0 & 0 & i h(\mu)   \\
0 & 0 & 0   \\
-i h(\mu) & 0 & 0   \\
	\end{array}
\right].\label{boundary-T3-3}
\end{align}
By component-wise, the matching conditions (\ref{matching}) for $N=3$ read from
(\ref{boundary-3}), (\ref{boundary-T1-3}), (\ref{boundary-T2-3}) and (\ref{boundary-T3-3}) as
\begin{align}
&\frac{1}{2}\mathrm{tr}_2\,\sigma_1\lambda^\dag\rho=-ig(\mu),&
&\frac{1}{2}\mathrm{tr}_2\,\sigma_1\lambda^\dag\chi=0,&
&\frac{1}{2}\mathrm{tr}_2\,\sigma_1\rho^\dag\chi=-ig(\mu),&\nonumber\\
&\frac{1}{2}\mathrm{tr}_2\,\sigma_2\lambda^\dag\rho=-ig(\mu),&
&\frac{1}{2}\mathrm{tr}_2\,\sigma_2\lambda^\dag\chi=0,&
&\frac{1}{2}\mathrm{tr}_2\,\sigma_2\rho^\dag\chi=ig(\mu),&\label{matching3}\\
&\frac{1}{2}\mathrm{tr}_2\,\sigma_3\lambda^\dag\rho=0,&
&\frac{1}{2}\mathrm{tr}_2\,\sigma_3\lambda^\dag\chi=i h(\mu),&
&\frac{1}{2}\mathrm{tr}_2\,\sigma_3\rho^\dag\chi=0,&\nonumber
\end{align}
By using the fact that the left hand side of (\ref{boundary-3}) is invariant 
under the multiplication of a unit quaternion $h$,
 \textit{i.e.}, $h^\dag h=1$, to $\tilde{W}$ from the left,
we can fix one of the quaternion component to be zero.
We choose the real component of $\lambda$, say, to be zero.
Thus, the remaining components of the boundary data to be determined are
\begin{equation}
\left.
\begin{array}{l}
\lambda=-i(\lambda_1\sigma_1+\lambda_2\sigma_2+\lambda_3\sigma_3),\\
\rho=\rho_0-i(\rho_1\sigma_1+\rho_2\sigma_2+\rho_3\sigma_3),\\
\chi=\chi_0-i(\chi_1\sigma_1+\chi_2\sigma_2+\chi_3\sigma_3).
\end{array}
\right.
\end{equation}
Then one can observe that a solution to (\ref{matching3}) is given in terms of 
 $\lambda_1,\lambda_2$ and $\lambda_3$, with $\lambda_1\neq\lambda_2$,
\begin{align}
&\rho_0=-\frac{\lambda_1+\lambda_2}{\Lambda^2}g(\mu),\quad
\rho_1=\rho_2=\frac{\lambda_3}{\Lambda^2}g(\mu),\quad 
\rho_3=\frac{\lambda_1-\lambda_2}{\Lambda^2}g(\mu)\nonumber\\
&\chi_0=\lambda_3,\quad  \chi_1=-\lambda_2,\quad \chi_2=\lambda_1,\quad\chi_3=0,
\label{general boundary data}
\end{align}
subject to a constraint 
\begin{align}
h(\mu)=\lambda_1^2+\lambda_2^2+\lambda_3^2=:\Lambda^2.\label{constraint:N=3}
\end{align}
There are two independent parameters in this boundary data due to the constraint (\ref{constraint:N=3}).
Note that the constraint (\ref{constraint:N=3}) gives a restriction $h(\mu)>0$ on the bulk data,
which reads
\begin{equation}
h(\mu)=-\frac{1}{2}\left(p_0(\mu)-p_1(\mu)\right)>0.
\end{equation}
For this condition to be enjoyed, we have to take an appropriate sign in the argument 
of theta function, such as $s_j=s+j/3$ for the $\vartheta_3$ solution, and
$s_j=-s+j/3$ for the $\vartheta_0$ solution, respectively.

Although the boundary data (\ref{general boundary data})  with (\ref{constraint:N=3}) is
a solution to (\ref{matching3}),  it does not enjoy the $C_3$-symmetry condition
\begin{equation}
R_3\otimes R_2\,W^\dag=W^\dag \hat q,\label{boundaryrotation3}
\end{equation}
in general,  
where $W=\tilde{W}U_3$, and $R_3$ and $R_2$ are the images of $C_3$-rotation.
To find the $C_3$-symmetric boundary data, 
we need to impose additional constraints $\lambda_1=\lambda_2$ and $\lambda_3=0$
on (\ref{general boundary data}) .
Then the solution (\ref{general boundary data}) reduces to
\begin{equation}
\begin{array}{l}
\lambda=i\lambda_1(\sigma_1+\sigma_2),\\
\rho=\rho_0=-\frac{2}{\lambda_1}g(\mu),\\
\chi=-i\lambda_1(\sigma_1-\sigma_2),
\end{array}\label{C3 boundary data}
\end{equation}
with a constraint $h(\mu)=2\lambda_1^2$, so that
there is no independent parameter.
We can observe this boundary data has the $C_3$-symmetry (\ref{boundaryrotation3})
around the ``$3$-axis",  with
\begin{equation}
R_2=\hat{q}=\cos\left(\frac{\pi}{3}\right)+i\sin\left(\frac{\pi}{3}\right)\,\sigma_3.
\end{equation}
Therefore, we have found the $C_3$-symmetric caloron Nahm data.

\subsection{Large scale and large period limit}
We now discuss some particular limits of the Nahm data.

Let us consider first the large scale, or monopole, limit of the caloron Nahm data.
If the caloron Nahm data has a monopole limit, the bulk data has to have a simple pole
at the boundary $s=\pm\mu$ whose residue belongs to an irreducible representation $\xi_j$
of $su(2)$ \cite{Hitchin},
\begin{equation}
\tilde{T}_j(s)\longrightarrow \frac{\xi_j}{s\mp\mu}+\cdots.
\end{equation}
Here it does not matter which basis of the bulk data are used.

All of the caloron Nahm data obtained in the former articles always
possess these monopole limits, because they are basically constructed 
by rearranging the Nahm data of monopoles \cite{LeeLu,KvB,BNvB,Harland,NakaSaka,Ward}.
In the case being considered here, the components of the bulk data do not have poles, or are 
non-singular on the region $[-\mu,\mu]$ for any value of $\mu$, 
found from (\ref{real T1})-(\ref{real T3}).
The value of the boundary data, or the scale parameters,
 of the caloron have some upper limits and non-divergent. 
In the reference \cite{SutNP}, the author constructed an approximated Nahm 
data of a monopole by the Laurent series expansion with a simple pole, which was 
sufficient to visualize the monopole energy.
He also claimed that if one would like to find the exact form of the monopole Nahm data, 
then one needed the Riemann theta function for surface of genus greater than one, which of course 
required much cumbersome efforts. 
For the analysis on the calorons, however, it is mandatory to know the analytical behavior 
of the Nahm data on the whole region $[-\mu,\mu]$ than only at the neighborhood of the boundary. 
We successfully constructed the Nahm data for the calorons, but they do 
not have large scale limits.

The other limit we consider is the large period, or the instanton, limit, which comes from by 
taking $\mu\to0$.
We restrict ourselves to the case with $C_3$-symmetry; 
For the case of the general boundary data (\ref{general boundary data}),
it is extremely complicated to analyze the large period limit.
In this limit, only the value of the bulk data at $s=0$ contributes, whose
explicit form is
\begin{align}
\tilde{T}_1(0)=
\frac{1}{2}\left[
\begin{array}{ccc}
0 & f_+^0 &f_0^0\\
f_+^0 & 0 &f_+^0\\
f_0^0 & f_+^0 & 0\\
\end{array}
\right],&\quad
\tilde{T}_2(0)=
\frac{1}{2}\left[
\begin{array}{ccc}
f_0^0 & -f_+^0 & 0\\
-f_+^0 & 0 & f_+^0\\
0 & f_+^0 & -f_0^0\\
\end{array}
\right],\nonumber\\
\tilde{T}_3(0)=
\frac{1}{4}&\left[
\begin{array}{ccc}
-p_2^0 & 0 & 0\\
0 & 2p_2^0 & 0\\
0 & 0 & -p_2^0\\
\end{array}
\right],\
\end{align}
where we defined $f_+^0:=f_+(0)$ \textit{etc.}.
The elements are given by the special values of the theta functions,
\begin{align}
f_+^0&:=f_+(0)=iC\frac{\sqrt{\vartheta_\nu(0)\vartheta_\nu(2\kappa)}}{\vartheta_\nu(\kappa)}
=iC\,\sqrt{\frac{\vartheta_\nu(0)}{\vartheta_\nu(\kappa)}},\label{fp0}\\
f_0^0&:=f_0(0)=iC\frac{\sqrt{\vartheta_\nu(-\kappa)\vartheta_\nu(\kappa)}}{\vartheta_\nu(0)}
=iC\,\frac{\vartheta_\nu(\kappa)}{\vartheta_\nu(0)},\label{f00}\\
p_2^0&:=p_2(0)=-\frac{\vartheta'_\nu(-\kappa)}{\vartheta_\nu(-\kappa)}+\frac{\vartheta'_\nu(\kappa)}{\vartheta_\nu(\kappa)}
=2\frac{\vartheta'_\nu(\kappa)}{\vartheta_\nu(\kappa)},\label{p20}
\end{align}
where $\kappa=1/3, \;\nu=0,3$ and we used the periodicity $\vartheta_\nu(2/3)=\vartheta_\nu(-1/3)$.
In the boundary data (\ref{C3 boundary data}), the contribution to the instanton limit comes from 
the lowest order terms in $\mu$.  
The $\mu$-dependence of the boundary data are fixed by the consistency with the matching conditions
(\ref{matching});
One can find the lowest order terms are $O(\sqrt{\mu})$,
\begin{align}   
\lambda_1=\sqrt{2\mu}\,\tilde{\lambda}_1+\dots,\label{lambda lowest}\\
\rho_0=-2\sqrt{2\mu}\,\tilde{\rho}_0+\dots,\label{rho lowest}
\end{align}
where 
\begin{align}
4\tilde{\lambda}_1^2
&=-C^2\left\{\left(\frac{\vartheta_\nu(\kappa)}{\vartheta_\nu(0)}\right)^2
-\frac{\vartheta_\nu(0)}{\vartheta_\nu(\kappa)}
\right\},\label{tildelambda}
\end{align}
and
\begin{equation}
\tilde{\rho}_0=\frac{iC}{4\tilde{\lambda}_1}\left(-3\vartheta'_\nu(\kappa)\right)
\sqrt{\frac{\vartheta_\nu(0)}{\vartheta^3_\nu(\kappa)}}.\label{tilderho}
\end{equation}
It can be found from (\ref{tildelambda}) that only the $\vartheta_0$ solution 
admits real $\tilde \lambda_1$; the right hand side is positive definite for 
pure imaginary $C$.
On the other hand, the right hand side is negative for the $\vartheta_3$ solution,
so that it is excluded from the large period limit.
This does not mean that the $\vartheta_3$ solution has no instanton limit; it only means that
the  $C_3$-symmetry and the instanton limit are not compatible.
If we consider the case of general boundary data, we expect that there exists the instanton limit of
the $\vartheta_3$ solution, since the behavior of the lowest order in $\mu$, 
(\ref{lambda lowest}) and (\ref{rho lowest}), must be modified.

Thus, we finally observe the Nahm data reduced to an ADHM matrix \cite{ADHM},
\begin{align}
&\Delta=\left[\begin{array}{ccc}
2i\tilde{\lambda}_1(\sigma_1+\sigma_2) & 2\tilde{\rho}_0  &-2i\tilde{\lambda}_1(\sigma_1-\sigma_2) \\
\\
&i\tilde{T}_j(0)\otimes\sigma_j&\\
\\
\end{array}\right]+
\left[
\begin{array}{ccc}
0& 0  &0  \\
x&0&0\\
0&x&0\\
0&0&x\\
\end{array}
\right],
\end{align}
where the $x$-independent part reads 
\begin{align}
\frac{1}{2}\left[
\begin{array}{ccc}
2i\tilde{\lambda}_1(\sigma_1+\sigma_2) & 2\tilde{\rho}_0  &-2i\tilde{\lambda}_1(\sigma_1-\sigma_2)  \\
{i}(f_0^0\sigma_2-\frac{1}{2}p_2^0\sigma_3)
& i f_+^0(\sigma_1-\sigma_2)
& i f_0^0\sigma_1\\
 i f_+^0(\sigma_1-\sigma_2)
& i p_2^0\sigma_3
& i f_+^0(-\sigma_1+\sigma_2)  \\
i f_0^0 \sigma_1 
& i f_+^0(\sigma_1+\sigma_2)
 & -i(f_0^0\sigma_2+\frac{1}{2}p_2^0\sigma_3)\\
\end{array}
\right].
\end{align}
One can show that the ADHM conditions for instantons
$\Im\Delta^\dag\Delta=0\Leftrightarrow \mathrm{tr}\sigma_j\Delta^\dag\Delta=0$
turn out to be
\begin{align}
4\tilde{\lambda}_1^2-(f_0^0)^2+(f_+^0)^2&=0\label{ADHM1}\\
\tilde{\lambda}_1\tilde{\rho}_0+\frac{3}{8}\,p_2^0\,f_+^0&=0,\label{ADHM2}
\end{align}
which are consistent with (\ref{tildelambda}) and (\ref{tilderho}) for the case of $\nu=0$ as expected.

\section{Conclusion}
In conclusion, we have considered the Nahm data of calorons with instanton charge $N$ which do not
have the large scale limit.
For the construction of the bulk Nahm data, we have applied the $C_N$-symmetric ansatz for the
monopoles given by .
As an illustration, the case of instanton charge $N=3$ is investigated in detail.
We have confirmed that there are $C_3$-symmetric calorons by a specific choice of the
boundary data, and given its instanton limit.
We  have also found the unitary transformation of the Nahm data between the $C_3$-symmetric basis
and the reality basis. 
To obtain the entire understanding on the $C_N$-symmetric calorons, 
it is necessary to investigate the case of larger $N$ in more detail,
 which will be given in the forthcoming articles.

From the perspective of the integrable systems,
it is significant to investigate the structure of ASD Yang-Mills solitons as a whole.
In this context, there are much works to do near future.
For example, it will be interesting to generalize the present work by introducing the theta functions of 
higher genus, the generalized 
Toda lattice corresponding to the affine Lie algebra rather than $\hat{A}^{(1)}_{N-1}$,
and so on.
In particular, it is quite attractive to consider that
 there are any other calorons without monopole limits.

\section*{Acknowledgement}
The authors would like to thank Shin Sasaki for his valuable comments.
They are also grateful to Satoru Saito for giving us the information on M.Toda's work.

\section*{Appendix}

In this appendix, we give the definition of the elliptic theta functions and their
properties necessary for the present article.
The elliptic theta functions are functions of two complex variables $u$ and $q$.
The latter is often called the modulus parameter.  
Introducing an auxiliary variable $z:=e^{u\pi i}$, they are defined by the 
Fourier series expansion in $u$,
\begin{align}
\vartheta_0(u,q)&=\sum_{n=-\infty}^\infty(-1)^nq^{n^2}z^{2n}=
1+2\sum_{n=1}^\infty(-1)^nq^{n^2}\cos2n\pi u, \\
\vartheta_1(u,q)&=i\sum_{n=-\infty}^\infty(-1)^nq^{(n-(1/2))^2}z^{2n}\nonumber\\
&=2\sum_{n=1}^\infty(-1)^{n-1}q^{(n-(1/2))^2}\sin(2n-1)\pi u,  \\
\vartheta_2(u,q)&=\sum_{n=-\infty}^\infty q^{(n-(1/2))^2}z^{2n-1} 
=2\sum_{n=1}^\infty q^{(n-(1/2))^2}\cos(2n-1)\pi u, \\
\vartheta_3(u,q)&=\sum_{n=-\infty}^\infty q^{n^2}z^{2n} =
1+2\sum_{n=1}^\infty q^{n^2}\cos2n\pi u. 
\end{align}
From these definitions, it is apparent that $\vartheta_1$ is odd function in $u$ 
and all of the others are even functions.
One can easily find their periodicity, $\vartheta_\nu(u+1)=\vartheta_\nu(u)$ for 
$\nu=0,3$ and $\vartheta_\nu(u+1)=-\vartheta_\nu(u)$ for $\nu=1,2$.
There are literatures in which the definition of the variable $u$ is scaled so that
the periodicity is $\vartheta_\nu(u+\pi)=\pm\vartheta_\nu(u)$.
It is also known that the elliptic theta functions have quasi-periodicity in
the complex $u$ plain, see for example \cite{WW}.

The elliptic theta functions enjoy many interesting relations.
Among them, the Jacobi triple product identities are well-known,
which gives infinite product representation of the Fourier series given above,
\begin{align}
\vartheta_0(u,q)&=q_0\prod_{n=1}^\infty(1-q^{2n-1}z^2)(1-q^{2n-1}/z^2)\nonumber\\&=
q_0\prod_{n=1}^\infty(1-2q^{2n-1}\cos2\pi u+q^{4n-2}), \\
\vartheta_1(u,q)&=-iq^{1/4}q_0z\prod_{n=1}^\infty (1-q^{2n}z^2)(1-q^{2n-2}/z^2)\nonumber\\
&=2q^{1/4}q_0\sin\pi u\prod_{n=1}^\infty (1-q^{2n}\cos2\pi u+q^{4n}),  \\
\vartheta_2(u,q)&=q^{1/4}q_0z\prod_{n=1}^\infty (1+q^{2n}z^2)(1+q^{2n-2}/z^2) \nonumber\\
&=2q^{1/4}q_0\cos\pi u\prod_{n=1}^\infty (1+q^{2n}\cos2\pi u+q^{4n}), \\
\vartheta_3(u,q)&=q_0z\prod_{n=1}^\infty (1+q^{2n-1}z^2)(1+q^{2n-1}/z^2) \nonumber\\
&=q_0\cos\pi u\prod_{n=1}^\infty (1+q^{2n-1}\cos2\pi u+q^{4n-2}),
\end{align}
where $q_0:=\prod_{n=1}^\infty(1-q^{2n})$.





\bibliographystyle{elsarticle-num}
\bibliography{<your-bib-database>}

\begin{thebibliography}{00}


\bibitem{ManSut} N.~Manton and P.~Sutcliffe, ``Topological Solitons",
Cambridge: Cambridge University Press (2004).
\bibitem{Radu} 
  E.~Radu and M.~S.~Volkov,
  Phys.\ Rept.\  {\bf 468} (2008) 101.

\bibitem{Chamon} 
  C.~Chamon, C.~-Y.~Hou, R.~Jackiw, C.~Mudry, S.~-Y.~Pi and A.~P.~Schnyder,
  Phys.\ Rev.\ Lett.\  {\bf 100} (2008) 110405.

\bibitem{Zahed} 
  I.~Zahed and G.~E.~Brown,
  Phys.\ Rept.\  {\bf 142} (1986)  1.

\bibitem{Shnir} 
  Ya.~M.~Shnir, ``Magnetic monopoles",
  Berlin: Springer (2005). 

\bibitem{AH} M.~Atiyah and N.~Hitchin
  ``The geometry and dynamics of matnetic monopoles",
  Princeton: Princeton university press (1988).


\bibitem{BPST} A.~A.~Belavin, A.~M.~Polyakov, A.~S.~Schwarz and Yu.~S.~Tyupkin,
Phys. Lett. B \textbf{59} (1975) 85.

\bibitem{Bogo} E.~B.~Bogomol'nyi, Sov. J. Nucl. Phys. \textbf{24} (1976) 449.
\bibitem{PS} M.~K.~Prasad and C.~M. Sommerfield, Phys. Rev. Lett. \textbf{35} (1975) 760.
\bibitem{HS} B.J.~Harrington and H.~K.~Shepard, Phys. Rev. D \textbf{17} (1978) 2122;
 Phys. Rev. D \textbf{18} (1978) 2990.
\bibitem{LeeLu}
  K.~-M.~Lee and C.~Lu,
  Phys. Rev. D {\bf 57} (1998) 5260;
\bibitem{KvB}
  T.~C.~Kraan and P.~van Baal,
  Phys.\ Lett.\ B {\bf 435} (1998) 389;
  Nucl.\ Phys.\ B {\bf 533} (1998) 627;
  Phys.\ Lett.\ B {\bf 428} (1998) 268.
\bibitem{BNvB}
  F.~Bruckmann, D.~Nogradi and P.~van Baal,
  Nucl.\ Phys.\ B {\bf 666} (2003) 197;
  Nucl.\ Phys.\ B {\bf 698} (2004) 233.
\bibitem{Harland}
  D.~Harland,
  J.\ Math.\ Phys.\  {\bf 48} (2007) 082905.
\bibitem{NakaSaka}
  A.~Nakamula and J.~Sakaguchi,
  J.\ Math.\ Phys.\  {\bf 51} (2010) 043503.


\bibitem{DiaGro} D.~Diaknov and N.~Gromov, Phys. Rev. D \textbf{72} (2005) 025003.
\bibitem{GroSli} N.~Gromov and S.~Slizovskiy, Phys. Rev. D \textbf{73} (2006) 025022.
\bibitem{Sli} S.~Slizovskiy, Phys. Rev. D \textbf{76} (2007) 085019.



\bibitem{SutPLB} P.~M.~Sutcliffe,
Phys. Lett. B \textbf{381} (1996) 129.
\bibitem{SutNP} P.~M.~Sutcliffe, Nucl. Phys. B \textbf{505} (1997) 517.






\bibitem{Br} H.~W.~Braden, 
Comm. Math. Phys. \textbf{308} (2011) 303.
\bibitem{HMM} N.~J.~Hitchin, N.~S.~Manton and M.~K.~Murray, 
Nonlinearity \textbf{8} (1995) 661.
\bibitem{ES} N.~Ercolani and A.~Sinha, Comm. Math. Phys. \textbf{125} (1989), 385-416.


\bibitem{Nahm} W.~Nahm,
Springer Lecture Notes in Phys. \textbf{201}(1984) 189. 
\bibitem{Ward} R.~S.~Ward, 
Phys. Lett. B \textbf{582}(2004) 203.


\bibitem{Toda} M.~Toda, ``Nonlinear waves and solitons", 
Tokyo: KTK Scientific Publishers/ Dordrecht, Boston, London: 
Kluwe Academic Publishers (1989)
\bibitem{Toda1} M.~Toda, 
Journ. Phys. Soc. Japan, \textbf{22} (1967), 431
\bibitem{WW} E.~T.~Whittaker and G.~N.~Watson, ``A course of mordern analysis",
Cambridge: Cambridge University Press (1902)


\bibitem{Hitchin} N.~J.~Hitchin, 
Comm. Math. Phys. \textbf{89} (1983) 145.




\bibitem{ADHM} M.~F.~Atiyah, V.~G.~Drinfeld, N.~J.~Hitchin and Yu.I.~Manin, 
Phys. Lett. A \textbf{65} (1978) 185.




\end{thebibliography}



\end{document}